\documentclass[12pt]{amsart}

\input{epsf}

\newtheorem{thm}{Theorem}[section]

\newtheorem{lem}[thm]{Lemma}

\theoremstyle{definition}

\theoremstyle{remark}

\numberwithin{equation}{section}

\begin{document}

\title[Unusual poles of the $\zeta$-functions ...]
{Unusual poles of the $\zeta$-functions for some regular singular
differential operators}
\author[H.\ Falomir, M.\ A.\ Muschietti,
P.\ A.\ G.\ Pisani and R.\ Seeley]{H.\ Falomir$^A$, M.\ A.\
Muschietti$^B$,
P.\ A.\ G.\ Pisani$^C$ and R.\ Seeley$^D$}%

\address{$A)$ IFLP, Departamento de F\'{\i}sica - Facultad de Ciencias
Exactas, UNLP, C.C. 67 (1900) La Plata, Argentina  \hfill\break
E-mail: falomir@fisica.unlp.edu.ar \ - \ Fax: (54 221) 425-2006}

\address{$B)$ Departamento de Matem\'{a}tica - Facultad de Ciencias Exactas,
UNLP, C.C. 172 (1900) La Plata, Argentina \hfill\break E-mail:
mariam@mate.unlp.edu.ar }

\address{$C)$ IFLP, Departamento de F\'{\i}sica - Facultad de Ciencias
Exactas, UNLP, C.C. 67 (1900) La Plata, Argentina \hfill\break
E-mail: pisani@fisica.unlp.edu.ar  }

\address{$D)$ University of Massachusetts at Boston, Boston, MA 02125,
USA \hfill\break E-mail: rts@math.umb.edu, r-seeley@attbi.com }



\date{August 12, 2003.}
\begin{abstract}
We consider the resolvent of a system of first order differential
operators with a regular singularity, admitting a family of
self-adjoint extensions. We find that the asymptotic expansion for
the resolvent in the general case presents powers of $\lambda$
which depend on the singularity, and can take even irrational
values. The consequences for the pole structure of the
corresponding $\zeta$ and $\eta$-functions are also discussed.

\bigskip

\noindent PACS numbers: 02.30.Tb, 02.30.Sa, 03.65.Db

\noindent Mathematical Subject Classification: 81Q10, 34L05, 34L40

\end{abstract}

\maketitle
\section{Introduction}

It is well known that in Quantum Field Theory under external
conditions, quantities like vacuum energies and effective actions,
which describe the influence of boundaries or external fields on
the physical system, are generically divergent and require a
renormalization to get a physical meaning.

In this context, a powerful and elegant regularization scheme to
deal with these problems is based on the use of the
$\zeta$-function \cite{Dowker,Hawking} or the heat-kernel (for
recent reviews see, for example,
\cite{Elizalde,Bytsenko,Klaus,Bordag,Dmitri}) associated to the
relevant differential operators appearing in the quadratic part of
the actions. In this way, ground state energies, heat-kernel
coefficients, functional determinants and partition functions for
quantum fields can be given in terms of the corresponding
$\zeta$-function, where the ultraviolet divergent pieces of the
one-loop contributions are encoded as poles of its holomorphic
extension.

Thus, it is of major interest in Physics to determine the
singularity structure of $\zeta$-functions associated with these
physical models.

\bigskip

In particular \cite{Seeley}, for an elliptic boundary value
problem in a $\nu$-dimensional compact manifold with boundary,
described by a differential operator $A$ of order $\omega$, with
smooth coefficients and a ray of minimal growth, defined on a
domain of functions subject to local boundary conditions, the
$\zeta$-function
\begin{equation}\label{zeta-func-def}
  \zeta_A(s)\equiv Tr\{A^{-s}\}
\end{equation}
has a meromorphic extension to the complex $s$-plane whose
singularities are isolated simple poles at $s=(\nu-j)/\omega$,
with $j=0,1,2,\dots$

In the case of positive definite operators, the $\zeta$-function
is related, via Mellin transform, to the trace of the heat-kernel
of the problem,  and the pole structure of $\zeta_A(s)$ determines
the small-$t$ asymptotic expansion of this trace
\cite{Seeley,Gilkey}:
\begin{equation}\label{heat-trace}
  Tr\{e^{-t A}\}\sim \sum_{j=0}^\infty a_j(A)\, t^{(j-\nu)/\omega},
\end{equation}
where the coefficients are related to the residues by
\begin{equation}\label{coef-res}
  a_j(A)=\left.{\rm Res}\right|_{s=(\nu-j)/\omega}
  \Gamma(s)\,\zeta_A(s).
\end{equation}

\bigskip

For operators $-(d/dx)^2 + V(x)$ with a singular potential $V(x)$
asymptotic to $g/ x^{2}$ as $x\rightarrow 0$, this expansion is
substantially different. If $g \geq 3/4$, the operator is
essentially self-adjoint. This case has been treated in
\cite{Callias1,Callias2,Callias3}, where log terms are found, as
well as terms with coefficients which are distributions
concentrated at the singular point $x=0$. For the case $g > -1/4$,
the Friedrichs extension has been treated in \cite{Bruening} for
operators in $\mathbf{L_2}(0,1)$, and in \cite{Bruening-Seeley}
for operators in $\mathbf{L_2}(\mathbf{R^+})$.

\bigskip

On the other hand, reference \cite{FPW} gave the pole structure of
the $\zeta$-function of a second order differential operator
defined on the (non compact) half-line $\mathbf{R}^+$,  having a
singular zero-th order term given by $V(x)=g \, x^{-2}+ x^2$. It
showed that, for a certain range of real values of $g$, this
operator admits nontrivial self-adjoint extensions in
$\mathbf{L_2}(\mathbf{R^+})$, for which the associated
$\zeta$-function presents isolated simple poles which (in general)
do not lie at $s=(1-j)/2$ for $j=0,1,\dots$, and can even take
irrational values.

A similar structure has been noticed in \cite{FP} for the
singularities of the $\zeta$-function of a system of first order
differential operators in the half line, appearing in a model of
Supersymmetric Quantum Mechanics with a singular superpotential
$\sim x^{-1}$.

\bigskip

Let us mention that singular potentials $\sim 1/x^2$ have been
considered in the description of several physical systems, like
the Calogero Model \cite{Calogero,Perelomov,FPW,Basu}, conformal
invariant quantum mechanical models \cite{DA-F-F,Camblong,Coon}
and, more recently, the dynamics of quantum particles in the
asymptotic  near-horizon region of black-holes
\cite{Claus,Gibbons,Govindarajan,Birmingham,Moretti}. Moreover,
singular superpotentials  has been considered as  possible agents
of supersymmetry breaking in models of Supersymmetric Quantum
Mechanics \cite{Jevicki,Pernice,Das} (see also \cite{FP}).

\bigskip

It is the aim of the present article to analyze the behavior of
the resolvent and $\zeta$ and $\eta$-functions of a system of
first order differential operators with a regular singularity in a
compact segment, admitting a family of self-adjoint extensions.

We will show that the asymptotic expansion for the resolvent in
the general case presents powers of $\lambda$ which depend on the
singularity, and can take even irrational values. The consequence
of this behavior on the corresponding $\zeta$ and $\eta$-functions
is the presence of simple poles lying at points which also depend
on the singularity, with residues depending on the self-adjoint
extension considered.

{We first construct the resolvents for two particular extensions,
for which the boundary condition at the singular point $x=0$ is
invariant under the scaling $x \rightarrow c\,x$. The resolvent
expansion for these special extensions displays the usual powers,
leading to the usual poles for the $\zeta$-function. The
resolvents of the remaining extensions are convex linear
combinations of these special extensions, but the coefficients in
the convex combination depend on the eigenvalue parameter
$\lambda$. This dependence leads to unusual powers in the
resolvent expansion, and hence to unusual poles for the
zeta-function. These self-adjoint extensions are not invariant
under the scaling $x\rightarrow c\,x$; as $c \rightarrow 0$ they
tend (at least formally) to one of the invariant extensions, and
as $c \rightarrow \infty$ they tend to the other. As $c
\rightarrow 0$ the residues at the anomalous poles tend to zero,
whereas as $c \rightarrow \infty$ these residues become infinite.
The way these residues depend on the boundary condition is
explained by a scaling argument in Section 7.}

\bigskip

The structure of the article is as follows: In Section
\ref{the-operator} we define the operator and determine its
self-adjoint extensions, and in Section \ref{the-spectrum} we
study their spectra. In Section \ref{the-resolvent} we construct
the resolvent for a general extension as a linear combination of
the resolvent of two limiting cases, and in Section
\ref{trace-resolvent} we consider the traces of these operators.
The asymptotic expansion of these traces, evaluated in Section
\ref{Asymptotic-expansion}, is used in Section
\ref{spectral-functions} to construct the associated $\zeta$ and
$\eta$-functions and study their singularities.

Finally, in Section \ref{second-order} we briefly describe similar
results one can obtain for a second order differential operator
with a regular singularity, also admitting a family of
self-adjoint extensions.

\section{The operator and its self-adjoint extensions}
\label{the-operator}

Let us consider the differential operator
\begin{equation}\label{D}
  D_x=\left(\begin{array}{cc}
    0  & \tilde{A}_x \\
    A_x & 0 \
  \end{array}\right)\, ,
\end{equation}
with
\begin{equation}\label{AA}
  A_x=-\partial_x + \frac{g}{x}=-
  x^{g} \, \partial_x\, x^{-g},
  \quad \tilde{A}_x = \partial_x + \frac{g}{x}=
  x^{-g}\,\partial_x \, x^{g}\, ,
\end{equation}
and $g\in \mathbb{R}$, defined on a domain of (two component)
smooth functions with compact support in a segment,
$\mathcal{D}(D_x)= \mathcal{C}_0^\infty(0,1)$. It can be easily
seen that $D_x$ so defined is symmetric.

The adjoint operator $D_x^*$, which is the maximal extension of
$D_x$, is defined on the domain $\mathcal{D}(D_x^*)$ of functions
$\Phi(x)= \left(\begin{array}{c}
  \phi_1(x) \\
  \phi_2(x)
\end{array}\right)\in \mathbf{L_2}(0,1)$, having a locally summable
first derivative and such that
\begin{equation}\label{DPhi}
    D_x\Phi(x)=\left( \begin{array}{c}
      \tilde{A}_x \phi_2(x) \\
      A_x \phi_1(x)
    \end{array} \right) = \left(\begin{array}{c}
      f_1(x) \\
      f_2(x)
    \end{array}\right)\in \mathbf{L_2}(0,1)\, .
\end{equation}

\bigskip

\begin{lem} \label{lema1-1}
If $\Phi(x)\in \mathcal{D}(D_x^*)$ and $-\frac{1}{2}< g<
\frac{1}{2}$, then
\begin{equation}\label{lemaI}
  \big|\,\phi_1(x)-C_1[\Phi]\, x^g\big| +
  \big|\,\phi_2(x)-C_2[\Phi]\, x^{-g}\big|
  \leq K_g\, \|D_x\Phi(x)\| \, x^{1/2}\, ,
\end{equation}
for some constants $C_1[\Phi]$ and $C_2[\Phi]$, where $\| \cdot
\|$ is the $\mathbf{L_2}$-norm.
\end{lem}

Indeed, Eqs.\ (\ref{DPhi}) and (\ref{AA}) imply
\begin{equation}\label{phi-chi-en0}
    \begin{array}{c}
      \phi_1(x)= C_1[\Phi]\, x^g - x^g \, \int_0^x y^{-g}\, f_2(y)\, dy
     \,  ,\\ \\
     \phi_2(x)= C_2[\Phi]\, x^{-g} + x^{-g} \, \int_0^x y^{g}\, f_1(y)\, dy
     \, ,
    \end{array}
\end{equation}
where $C_1[\Phi]$ and $C_2[\Phi]$ are integration constants which
depend on the function $\Phi(x)$. Taking into account that
\begin{equation}\label{schwarz}
    \begin{array}{c}\displaystyle{
      \left|\int_0^x y^{g} \, f_1(y)\, dy\right|\leq
      \frac{x^{g+1/2}}{\sqrt{1+2g}}\, \|f_1\| }\, ,
      \\ \\ \displaystyle{
      \left|\int_0^x y^{-g} \, f_2(y)\, dy\right|\leq
      \frac{x^{-g+1/2}}{\sqrt{1-2g}}\, \|f_2\| }\,  ,
    \end{array}
\end{equation}
we immediately get Eq.\ (\ref{lemaI}) with $K_g =
(1-2g)^{-1/2}+(1+2g)^{-1/2}$.

\bigskip

\begin{lem}
Let $\Phi(x)=\left(\begin{array}{c}
  \phi_1(x) \\
  \phi_2(x)
\end{array}\right),\Psi(x)=\left(\begin{array}{c}
  \psi_1(x) \\
  \psi_2(x)
\end{array}\right) \in \mathcal{D}(D^*)$. Then
\begin{equation}\label{DDstar}\begin{array}{c}
    \left(D_x \Psi, \Phi\right) -
  \left(\Psi, D_x \Phi\right) = \\ \\
  = \Big\{
  C_1[\Psi]^* C_2[\Phi] - C_2[\Psi]^* C_1[\Phi]\Big\}+
  \Big\{\psi_2(1)^*\, \phi_1(1)
  - \psi_1(1)^*\, \phi_2(1) \Big\}\, .
\end{array}
\end{equation}

\end{lem}

In fact, from Eq.\ (\ref{AA}) one easily obtains
\begin{equation}\label{DDstar2}\begin{array}{c}
      \left(D_x \Psi, \Phi\right) -
  \left(\Psi, D_x \Phi\right) = \\ \\
    =\displaystyle{
    \lim_{\varepsilon \rightarrow 0^+}\int_\varepsilon^1
    \partial_x \Big\{x^g\,\psi_2(x)^*\, x^{-g}\,\phi_1(x)
    - x^{-g}\,\psi_1(x)^* \, x^g \, \phi_2(x)
    \Big\} dx}\, ,
\end{array}
\end{equation}
from which, taking into account the results in Lemma
\ref{lema1-1}, Eq.\ (\ref{DDstar}) follows directly.

\bigskip

Now, if $\Psi(x)$ in Eq.\ (\ref{DDstar}) belongs to the domain of
the closure of $D_x$, $\overline{D}_x=(D_x^{*})^*$,
\begin{equation}\label{en-la-clausura}
  \Psi(x)\in
\mathcal{D}(\overline{D}_x) \subset \mathcal{D}(D_x^*)\,  ,
\end{equation}
then the right hand side of Eq.\ (\ref{DDstar}) must vanish for
any $\Phi(x)\in \mathcal{D}(D_x^*)$. Therefore
\begin{equation}\label{Psi-clausura}
    C_1[\Psi]=0=C_2[\Psi]\, ,\ {\rm and}\  \Psi(1)=0\, .
\end{equation}

\bigskip

On the other hand, if $\Psi(x),\Phi(x)$ belong to the domain of a
symmetric extension of $D_x$ (contained in $\mathcal{D}(D_x^*)$),
the right hand side of Eq.\ (\ref{DDstar}) must also vanish.

Thus, the closed extensions of $D_x$ correspond to the subspaces
of $\mathbb{C}^4$ under the map $\Phi\rightarrow \left( C_1[\Phi],
C_2[\Phi], \phi_1(1), \phi_2(1) \right)$, and the self-adjoint
extensions correspond to those subspaces $S\subset \mathbb{C}^4$
such that $S=S^{\perp}$, with the orthogonal complement taken in
the sense of the symplectic  form on the right hand side of Eq.\
(\ref{DDstar}).

\bigskip

For definiteness, in the following we will consider self-adjoint
extensions satisfying the local boundary condition
\begin{equation}\label{BC1}
  \phi_1(1)=0\, .
\end{equation}
Each such extension is determined by a condition of the form
\begin{equation}\label{BC2}
    \alpha\, C_1[\Phi] + \beta\, C_2[\Phi] = 0\, ,
\end{equation}
with $\alpha,\beta \in \mathbb{R}$, and $\alpha^2 + \beta^2 \neq 0
$. We denote this extension by $D_x^{(\alpha, \beta)}$.

\section{The spectrum} \label{the-spectrum}

In order to determine the spectrum of the self-adjoint extensions
of $D_x$, we need the solutions of
\begin{equation}\label{Ec-hom}
  (D_x-\lambda)\Phi(x)=0 \Rightarrow \left\{
  \begin{array}{c}
    \tilde{A}_x \phi_2(x) = \lambda \phi_1(x)\, ,\\ \\
    A_x \phi_1(x) = \lambda \phi_2(x)\, ,
  \end{array}\right.
\end{equation}
satisfying the boundary conditions in Eqs.\ (\ref{BC1}) and
(\ref{BC2}).

\bigskip

The solution of the homogeneous equation for $\lambda=0$  is
\begin{equation}\label{lambda0}
  \Phi(x)=
\begin{pmatrix}
  C_1 \, x^g\\
  C_2 \, x^{-g}
\end{pmatrix}\, ,
\end{equation}
but the boundary conditions in Eqs.\ (\ref{BC1}) and (\ref{BC2})
imply that $C_1 = 0$ and $C_2=0$, unless  $\beta = 0$.
Consequently, there are no zero modes except for the self-adjoint
extension characterized by $\beta =0$, $D_x^{(1,0)}$.

\bigskip

Applying $\tilde{A}$ to the second line in Eq.\ (\ref{Ec-hom}),
and using the first one, one easily gets
\begin{equation}\label{hom-sec}
    \left\{ \partial_x^2 - \frac{g(g-1)}{x^2}
    + \lambda^2\right\} \phi_1(x)=0\, .
\end{equation}
Then, for $\lambda\neq 0$, the solutions are of the form
\begin{equation}\label{sol-hom-1}
  \phi_1(x) = K_1\, \sqrt{X}\, J_{\frac 1 2 -g}(X)
  + K_2 \, \sqrt{X} \, J_{g-\frac 1 2}(X)\, ,
\end{equation}
with $X= \tilde{\lambda} \, x$, where
$\tilde{\lambda}=+\sqrt{\lambda^2}$ and $K_1, K_2$ are constants.

This implies for the lower component of $\Phi(x)$
\begin{equation}\label{hom-2}
  \phi_2(x)= \sigma
  \left\{-K_1 \, \sqrt{X}\, J_{-g-\frac 1 2}(X)
  + K_2\, \sqrt{X} \, J_{g+\frac 1 2}(X)\right\}\, ,
\end{equation}
where $\sigma={\tilde{\lambda}}/{\lambda}$.

Taking into account that
\begin{equation}\label{J-Bessel}
  J_\nu(X)= X^{\nu }\,\left\{ \frac{1}
     {2^{\nu }\,
       \Gamma(1 + \nu )}
     + O(X^2) \right\}\, ,
\end{equation}
we get
\begin{equation}\label{ec-spectrum}\begin{array}{c}
    \alpha\, C_1[\Phi] + \beta\, C_2[\Phi] = \\ \\
    \displaystyle{ =
    \frac{\alpha\, K_2\, \tilde{\lambda}^g}
    {2^{g-\frac{1}{2}}
  \Gamma\left(\frac{1}{2}+g\right)}
  - \sigma
  \frac{\beta\, K_1\,
  \tilde{\lambda}^{-g}}{2^{-g-\frac{1}{2}}
  \Gamma\left(\frac{1}{2}-g\right)}=0}\, .
\end{array}
\end{equation}

\bigskip

For $\alpha = 0$, Eq.\ (\ref{ec-spectrum}) implies $K_1 = 0$.
Therefore, $\phi_1(1)=0 \Rightarrow  J_{g-\frac 1
2}(\tilde{\lambda})=0$. Thus, the spectrum of this extension,
$D_x^{(0,1)}$, is non-degenerate and symmetric with respect to the
origin, with the eigenvalues given by
\begin{equation}\label{alpha0}
  \lambda_{\pm,n} = \pm j_{g-\frac 1 2,n}\, ,
   \quad n=1,2,\dots\, ,
\end{equation}
where $j_{\nu,n}$ is the $n$-th positive zero of the Bessel
function $J_{\nu}(z)$\footnote{\label{zeroes} Let us recall that
large zeros of $J_{\nu}(\lambda)$ have the asymptotic expansion
\begin{equation}\label{large-zeroes}
  j_{\nu,n}\simeq \gamma -\frac{4 \nu^2-1}{8 \gamma}+O\left(\frac 1
  \gamma\right)^3\, ,
\end{equation}
with $\gamma=\left( n+\frac{\nu}{2}-\frac{1}{4} \right) \pi$.}.

\bigskip

For $\alpha\neq 0$, from Eq.\ (\ref{ec-spectrum}) we can write
\begin{equation}\label{spectrum}
  \frac{K_2}{K_1} = \sigma\,
  \tilde{\lambda}^{-2 g} \left[
  \frac{4^{g}\, \Gamma\left(\frac{1}{2}+g\right)}
  {\Gamma\left(\frac{1}{2}-g\right)}\right]
  \left(\frac{\beta}{\alpha}\right)\, .
\end{equation}
In this case, the boundary condition at $x=1$ determines the
eigenvalues as the solutions of the transcendental equation
\begin{equation}\label{eigenvalues}
  \tilde{\lambda}^{2 g}\,
  \frac{J_{\frac 1 2 -g}(\tilde{\lambda})}
  {J_{g-\frac 1 2}(\tilde{\lambda})}
  =\sigma \, \rho(\alpha,\beta)\, ,
\end{equation}
where we have defined
\begin{equation}\label{rho}
  \rho(\alpha,\beta):=
   -\frac{4^{g}\, \Gamma\left(\frac{1}{2}+g\right)}
  {\Gamma\left(\frac{1}{2}-g\right)}\,
  \left(\frac{\beta}{\alpha}\right)\, .
\end{equation}

\begin{figure}
    \epsffile{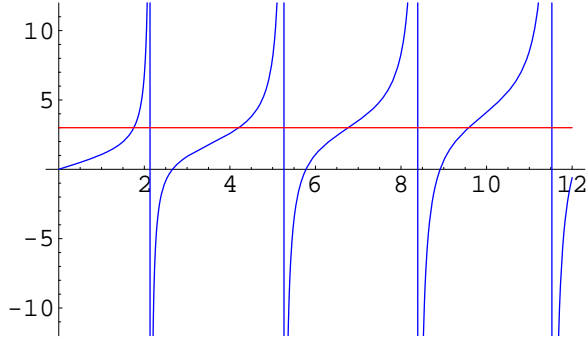}
    \caption{Plot for $F(\lambda):= \displaystyle{\lambda^{2 g}\,
  \frac{J_{\frac 1 2 -g}(\lambda)}{J_{g-\frac 1 2}(\lambda)}
  }$, with $g=1/3$, and $\rho(\alpha,\beta)=3$.} \label{figure}
\end{figure}

For the positive eigenvalues $\tilde{\lambda}=\lambda\Rightarrow
\sigma =1 $, and Eq.\ (\ref{eigenvalues}) reduces to
\begin{equation}\label{eigenvalues-pos}
  F(\lambda):=\lambda^{2 g}\,
  \frac{J_{\frac 1 2 -g}(\lambda)}{J_{g-\frac 1 2}(\lambda)}
  = \rho(\alpha,\beta)\, ,
\end{equation}
relation plotted in Figure \ref{figure} for particular values of
$\rho(\alpha,\beta)$ and $g$.

On the other hand, for negative eigenvalues $\lambda = e^{i\, \pi}
\tilde{\lambda}\Rightarrow \sigma = \tilde{\lambda}/\lambda =
e^{-i\, \pi}$, and Eq.\ (\ref{eigenvalues}) reads as
\begin{equation}\label{eigenvalues-neg}
  F(\tilde{\lambda})= e^{-i\, \pi} \rho(\alpha,\beta)
  =\rho(\alpha,-\beta)\, .
\end{equation}
Therefore, the negative eigenvalues of $D_x^{(\alpha, \beta)}$ are
$e^{i\, \pi}$ times the positive eigenvalues of $D_x^{(\alpha,
-\beta)}$.

\bigskip

Notice that the spectrum is always non-degenerate, and there is a
positive eigenvalue between each pair of consecutive zeroes of
$J_{g-\frac 1 2}(\lambda)$.

Moreover, the spectrum is symmetric with respect to the origin
only for the $\alpha=0$ extension (which we call the
``D-extension", see Eq.\ (\ref{alpha0})), and for the $\beta=0$
extension (which we call the ``N-extension"). Indeed, in this last
case, from Eqs.\ (\ref{eigenvalues}) and (\ref{rho}) one can see
that the eigenvalues of $D_x^{(1,0)}$ are given by
\begin{equation}\label{eigen-beta0}
  \lambda_0 =0\, , \quad
  \lambda_{\pm,n} = \pm j_{\frac 1 2 -g ,n}\, , \ n=1,2,\dots
\end{equation}

\section{The resolvent} \label{the-resolvent}

In this Section we will construct the resolvent of $D_x$,
\begin{equation}\label{def-resolv}
  G(\lambda)=(D_x-\lambda)^{-1}\, ,
\end{equation}
for its different self-adjoint extensions.

We will first consider the two limiting cases in Eq.\ (\ref{BC2}),
namely the ``$D$-extension", for which $\alpha=0 \Rightarrow
C_2[\Phi] =0$, and the ``$N$-extension", with $\beta=0 \Rightarrow
C_1[\Phi]=0$. The resolvent for a general self-adjoint extension
will be later evaluated as a linear combination of those obtained
for these two limiting cases.

\bigskip

For the kernel of the resolvent 
\begin{equation}\label{resolv}
  G(x,y; \lambda)=\left(\begin{array}{cc}
    G_{11}(x,y; \lambda) & G_{12}(x,y; \lambda) \\
    G_{21}(x,y; \lambda) & G_{22}(x,y; \lambda)
  \end{array}\right)\, ,
\end{equation}
we have
\begin{equation}\label{G}
   (D_x-\lambda)\, G(x,y; \lambda) =
\delta(x,y)\, \mathbf{1}_2\, ,
\end{equation}
from which we straightforwardly  get for the diagonal elements
\begin{equation}\label{ec-dif-g-diag}
  \begin{array}{c}
  \displaystyle{
    \left\{ \partial_x^2 - \frac{g(g-1)}{x^2}
    + \lambda^2 \right\}  G_{11}(x,y; \lambda) = -\lambda\,
     \delta(x,y)\, ,} \\\\
       \displaystyle{
       \left\{ \partial_x^2 - \frac{g(g+1)}{x^2}
    + \lambda^2 \right\}  G_{22}(x,y; \lambda) = -\lambda\,
     \delta(x,y) }\, ,
  \end{array}
\end{equation}
while for the non diagonal ones we have
\begin{equation}\label{ec-dif-g-nodiag}
  \begin{array}{c}
      \displaystyle{
    G_{21}(x,y; \lambda)= \frac{1}{\lambda}
    \left\{- \partial_x + \frac g x \right\}
    G_{11}(x,y; \lambda)\, , }\\ \\
      \displaystyle{
    G_{12}(x,y; \lambda)= \frac{1}{\lambda}
    \left\{ \partial_x + \frac g x \right\}
    G_{22}(x,y; \lambda)\, ,}
  \end{array}
\end{equation}
for $\lambda\neq 0$.

Since the resolvent is analytic in $\lambda$, it is sufficient to
evaluate it on the open right half plane.

\bigskip

In so doing,  we will need the upper and lower components of some
particular solutions of the homogeneous equation (\ref{Ec-hom}).

Then, let us define
\begin{equation}\label{soluciones}
    \left\{
  \begin{array}{l}
    L_1^D(X)= \sqrt{X}\, J_{g-\frac 1 2}(X)\,  ,\\ \\
        L_2^D(X)= \sqrt{X}\, J_{g+\frac 1 2}(X)\,  ,\\ \\
        L_1^N(X)= \sqrt{X}\, J_{\frac 1 2 -g}(X)\,  ,\\ \\
        L_2^N(X)= \sqrt{X}\, J_{-g-\frac 1 2}(X)\, , \\ \\
    R_1(X;\lambda)=\sqrt{X} \left[J_{g-\frac 1 2}({\lambda})
     J_{\frac 1 2 -g}(X)
    - J_{\frac 1 2 -g}({\lambda}) J_{g-\frac 1 2}(X)\right]\, ,
    \\ \\
    R_2(X;\lambda)=\sqrt{X} \left[J_{g-\frac 1 2}({\lambda})
    J_{-g-\frac 1 2}(X)
    + J_{\frac 1 2 -g}({\lambda}) J_{g+\frac 1 2}(X)\right]\, .
  \end{array}
  \right.
\end{equation}
Notice that $R_1({\lambda};\lambda)=0$, and $\left.\tilde{A_x}\,
R_2({\lambda}\, x;\lambda )\right|_{x=1}=0$.

We will also need the Wronskians
\begin{equation}\label{W-D}
\left\{
\begin{array}{c}
  \displaystyle{
  W\left[ L_1^D(X), R_1(X;\lambda) \right] =
   - \frac 2 \pi \, \cos(g\, \pi) \,
  J_{g-\frac 1 2}({\lambda})=\frac{1}{ \gamma_D(\lambda)}\, ,
    } \\ \\
  \displaystyle{
  W\left[ L_2^D(X), R_2(X;\lambda) \right] =
   \frac 2 \pi \, \cos(g\, \pi) \,
  J_{g-\frac 1 2}({\lambda})=\frac{-1}{ \gamma_D(\lambda)}\, ,
  } \\ \\
    \displaystyle{
  W\left[ L_1^N(X), R_1(X;\lambda) \right] =
   - \frac 2 \pi \, \cos(g\, \pi) \,
  J_{\frac 1 2 -g}({\lambda})=\frac{1}{ \gamma_N(\lambda)}\, ,
    } \\ \\
  \displaystyle{
  W\left[ L_2^N(X), R_2(X;\lambda) \right] =
   - \frac 2 \pi \, \cos(g\, \pi) \,
  J_{\frac 1 2-g}({\lambda})=\frac{1}{ \gamma_N(\lambda)}
  }\, ,
  \end{array}
  \right.
\end{equation}
which  vanish only at the zeroes of $J_{\nu}({\lambda})$, for $\nu
= \pm\left(\frac 1 2 -g\right)$.

\bigskip

\subsection{The resolvent for the $D$-extension} In this case,
the function
\begin{equation}\label{func-inhom}
  \Phi(x) = \int_0^1 G_D( x, y; \lambda)
  \left( \begin{array}{c}
    f_1(y) \\
    f_2(y)
  \end{array} \right) dy
\end{equation}
must satisfy $\phi_1(1)=0$ and $C_2[\Phi]=0$, for any functions
$f_1(x),f_2(x)\in \mathbf{L_2}(0,1)$.

This requires that
\begin{equation}\label{GD11}
  G_{11}^D (x,y; \lambda)= \gamma_D(\lambda) \times  \left\{
  \begin{array}{c}
    L_1^D(X)\, R_1(Y; \lambda),\ {\rm for}\ x\leq y\, , \\ \\
    R_1(X; \lambda)\, L_1^D(Y), \ {\rm for}\ x \geq y\, ,
  \end{array}\right.
\end{equation}
and
\begin{equation}\label{GD22}
  G_{22}^D (x,y; \lambda)= - \gamma_D(\lambda) \times  \left\{
  \begin{array}{c}
    L_2^D(X)\, R_2(Y; \lambda),\ {\rm for}\ x\leq y\, , \\ \\
    R_2(X; \lambda)\, L_2^D(Y), \ {\rm for}\ x \geq y\, ,
  \end{array}\right.
\end{equation}
with the other components, $G_{12}^D(x,y; \lambda)$ and
$G_{21}^D(x,y; \lambda)$, given as in Eq.\
(\ref{ec-dif-g-nodiag}). The fact that the boundary conditions are
satisfied, as well as $(D_x-\lambda)\, \Phi(x)=\begin{pmatrix}
  f_1(x) \\
  f_2(x)
\end{pmatrix}$, can be straightforwardly verified from Eqs.\
(\ref{soluciones} - \ref{W-D}).

\bigskip

Indeed, from Eqs.\ (\ref{func-inhom} - \ref{GD22}),
(\ref{ec-dif-g-nodiag}) and (\ref{soluciones} - \ref{W-D}), one
gets
\begin{equation}\label{near0-D}
    \phi_1(x)=
    C_1^D[\Phi] \, x^g + O(\sqrt{x})\, ,\quad
    \phi_2(x)= O(\sqrt{x})\,  ,
\end{equation}
with
\begin{equation}\label{C+}\begin{array}{c}
  \displaystyle{C_1^D[\Phi] =   \frac{-\, \pi\, \lambda^{g+1}}
    {2^{\frac{1}{2}+ g} \cos(g\, \pi)
    J_{g-\frac{1}{2}}({\lambda}) \,
    \Gamma\left(\frac{1}{2} +g\right)}}
    \times  \\ \\
  \displaystyle{  \times
    \int_0^1 \Big[R_1(\lambda\, y; \lambda) f_1(y)
    - R_2(\lambda\, y; \lambda) f_2(y)\Big] dy}\, ,
\end{array}
\end{equation}
for $\lambda$ not a zero of $J_{g-\frac{1}{2}}({\lambda})$.

\bigskip

Notice that $C_1^D[\Phi]\neq 0$ if the integral in the right hand
side of Eq.\ (\ref{C+}) is non vanishing.

\bigskip

\subsection{The resolvent for the $N$-extension} In this case,
the function
\begin{equation}\label{func-inhom-N}
  \Phi(x) = \int_0^1 G_N(x, y; \lambda)
  \left( \begin{array}{c}
    f_1(y) \\
    f_2(y)
  \end{array} \right) dy
\end{equation}
must satisfy $\phi_1(1)=0$ and $C_1[\Phi]=0$, for any functions
$f_1(x),f_2(x)\in \mathbf{L_2}(0,1)$.

This requires that
\begin{equation}\label{GN11}
  G_{11}^N (x,y; \lambda)= \gamma_N(\lambda) \times  \left\{
  \begin{array}{c}
    L_1^N(X)\, R_1(Y; \lambda),\ {\rm for}\ x\leq y\, , \\ \\
    R_1(X; \lambda)\, L_1^N(Y), \ {\rm for}\ x \geq y\, ,
  \end{array}\right.
\end{equation}
and
\begin{equation}\label{GN22}
  G_{22}^N (x,y; \lambda)= \gamma_N(\lambda) \times  \left\{
  \begin{array}{c}
    L_2^N(X)\, R_2(Y; \lambda),\ {\rm for}\ x\leq y\, , \\ \\
    R_2(X; \lambda)\, L_2^N(Y), \ {\rm for}\ x \geq y\, ,
  \end{array}\right.
\end{equation}
with the other components, $G_{12}^N(x,y; \lambda)$ and
$G_{21}^N(x,y; \lambda)$, given as in Eq.\
(\ref{ec-dif-g-nodiag}). These boundary conditions, as well as the
fact that $(D_x-\lambda)\, \Phi(x)=\begin{pmatrix}
  f_1(x) \\
  f_2(x)
\end{pmatrix}$, can be
straightforwardly verified from Eq.\ (\ref{soluciones} -
\ref{W-D}).

\bigskip

This time, from Eqs.\ (\ref{func-inhom-N} - \ref{GN22}),
(\ref{ec-dif-g-nodiag}) and (\ref{soluciones} - \ref{W-D}), one
gets
\begin{equation}\label{near0-N}
    \phi_1(x)=
    O(\sqrt{x})\, ,\quad
    \phi_2(x)= C_2^N[\Phi] \, x^{-g} + O(\sqrt{x})\, ,
\end{equation}
with
\begin{equation}\label{C-}\begin{array}{c}
  \displaystyle{C_2^N[\Phi] =   \frac{\pi\, \lambda^{1-g}}
    {2^{\frac{1}{2}- g} \cos(g\, \pi)
    J_{\frac{1}{2}-g}({\lambda}) \,
    \Gamma\left(\frac{1}{2}-g\right)} \times} \\ \\
  \displaystyle{ \phantom{C_2[\Phi] = } \times
    \int_0^1 \Big[R_1(\lambda\, y; \lambda) f_1(y)
    - R_2(\lambda\, y; \lambda) f_2(y)\Big] dy\, ,}
\end{array}
\end{equation}
for $\lambda$ not a zero of $J_{\frac{1}{2}-g}({\lambda})$.

\bigskip

Notice that $C_2^N[\Phi]\neq 0$ if the integral in the right hand
side of Eq.\ (\ref{C-}) (the same integral as the one appearing in
the $D$-extension, Eq.\ (\ref{C+})) is non vanishing.

\bigskip

\subsection{The resolvent for a general self-adjoint extension of $D_x$}

For the general case, we can adjust the boundary conditions
\begin{equation}\label{BC-general}
  \phi_1(1)=0\, ,\quad \alpha\, C_1[\Phi] + \beta\, C_2[\Phi] =
  0\, ,\ \alpha,\beta\neq 0\, ,
\end{equation}
for
\begin{equation}\label{func-inhom-gen}
  \Phi(x) = \int_0^1 G( x,  y; \lambda)
  \left( \begin{array}{c}
    f_1(y) \\
    f_2(y)
  \end{array} \right) dy\, ,
\end{equation}
for any $f_1(x),f_2(x)\in \mathbf{L_2}(0,1)$, by taking a linear
combination of the resolvent for the limiting cases,
\begin{equation}\label{linear-comb}
  G(x,y; \lambda)= \left[1- \tau(\lambda)\right] G_D(x,y; \lambda) +
  \tau(\lambda)\, G_N(x,y; \lambda)\, .
\end{equation}

Since the boundary condition at $x=1$ is automatically fulfilled,
one must just impose
\begin{equation}\label{ec-tau}
  \alpha \left[1- \tau(\lambda)\right] C_1^D[\Phi]
  + \beta\, \tau(\lambda)\, C_2^N[\Phi] =0\, .
\end{equation}

Notice that, in view of Eq.\ (\ref{C+}), (\ref{C-}) and
(\ref{eigenvalues-pos}),
\begin{equation}\label{nocero}
  \alpha \, C_1^D[\Phi]-\beta\,  C_2^N[\Phi]=0
\end{equation}
precisely when $\lambda$ is an eigenvalue of
$D_x^{(\alpha,\beta)}$. Therefore, from Eq.\ (\ref{ec-tau}) we get
the resolvent of $D_x^{(\alpha,\beta)}$ by setting
\begin{equation}\label{taudelambda}\begin{array}{c}
    \tau(\lambda) = \displaystyle{\frac{\alpha \, C_1^D[\Phi]}
  {\alpha \, C_1^D[\Phi]-\beta\,  C_2^N[\Phi]}} =
     \frac{1}{1-\displaystyle{
        \frac{\rho(\alpha,\beta)}{F(\lambda)}}} =
  \\  \\ =
  1-\displaystyle{ \frac{1}
  {1 - \displaystyle{\frac{{\lambda }^{2\,g}}{\rho(\alpha,\beta)}\
  \frac{ J_{\frac{1}
          {2} - g}(\lambda )}{
       J_{g-\frac{1}{2}}(\lambda )} }} \, ,
          }
\end{array}
\end{equation}
for  $\lambda$ not a zero of $J_{g-\frac{1}{2}}(\lambda )$.

\bigskip

\section{The trace of the resolvent} \label{trace-resolvent}

It follows from Eq.\ (\ref{linear-comb}) that the resolvent of a
general self-adjoint extension of $D_x$ can be expressed in terms
of the resolvents of the two limiting cases, $G_D(\lambda)$ and
$G_N(\lambda)$. Moreover, since the eigenvalues of any extension
grow linearly with $n$ (see Section \ref{the-spectrum}), these
resolvents are Hilbert-Schmidt operators and their
$\lambda$-derivatives are trace class.

So, let us consider the relation
\begin{equation}\label{derivG}\begin{array}{c}
    G^2(\lambda)=\partial_\lambda G(\lambda) =
    \partial_\lambda G_D(\lambda) - \\ \\
  - \tau'(\lambda) \left[G_D(\lambda)-G_N(\lambda)\right]
  - \tau(\lambda)
  \left[ \partial_\lambda G_D(\lambda) -
  \partial_\lambda G_N(\lambda) \right]\, ,
\end{array}
\end{equation}
from which it follows that the difference
$G_D(\lambda)-G_N(\lambda)$ is a strongly analytic function of
$\lambda$ (except at the zeroes of $\tau'(\lambda)$) taking values
in the trace class operators ideal.

\bigskip

Since we have explicitly constructed $G_D(\lambda)$ and
$G_N(\lambda)$ in the previous Section (see Eqs.\ (\ref{GD11}),
(\ref{GD22}), (\ref{GN11}) and (\ref{GN22})), we straightforwardly
get (see Appendix \ref{integrals} for the details)
\begin{equation}\label{traza-derivGD}\begin{array}{c}
        Tr\{\partial_\lambda G_D(\lambda)\} = \displaystyle{
      \int_0^1
      tr\{
      \partial_\lambda G_D( x, x; \lambda)\}\, dx} =
    \\ \\
    = \displaystyle{  \partial_\lambda
      \left\{\frac{J_{g+ \frac 1 2 }(\lambda)}
      {J_{g-\frac 1 2}(\lambda)}\right\} =
      1-\frac{2 g}{\lambda}\,
    \frac{J_{g+ \frac 1 2 }(\lambda)}{J_{g-\frac 1 2}(\lambda)}
    +\frac{J_{g+ \frac 1 2 }^2(\lambda)}{J_{g-\frac 1
    2}^2(\lambda)}
    =} \\ \\
    \displaystyle{
    =1 - \frac{g^2}{{\lambda }^2} +
  {\left( \frac{1}{2\,\lambda } +
      \frac{{J'_{g- \frac{1}{2} }}( \lambda )}
      {J_{g- \frac{1}{2}}(\lambda )} \right) }^2 \, ,}
\end{array}
\end{equation}
where, in the last step, we have taken into account that
\begin{equation}\label{numasmenos1}
  J_{\nu\pm 1}(z) = \frac \nu z \, J_\nu(z)
  \mp J'_\nu(z)\, .
\end{equation}
Similarly,
\begin{equation}\label{TrGD-GN}\begin{array}{c}
      Tr\{G_D(\lambda)-G_N(\lambda)\}
    =\displaystyle{
    \frac{2 g}{\lambda} +
        \frac{J_{g+\frac 1 2  }(\lambda)}{J_{g-\frac 1
        2}(\lambda)} +
    \frac{J_{-g - \frac 1 2}(\lambda)}
    {J_{-g+\frac 1 2}(\lambda)}=
    } \\ \\
  \displaystyle{= -\frac{2\,g}{\lambda } -
  \frac{{J'_{\frac{1}{2} - g}}(\lambda )}
  {J_{\frac{1}{2} - g}(\lambda )} +
  \frac{{J'_{g- \frac{1}{2}}}(\lambda )}
  {J_{g-\frac{1}{2}}(\lambda )}}\, .
\end{array}
\end{equation}

Moreover, since
\begin{equation}\label{derTr}
  \partial_\lambda Tr\{G_D(\lambda)-G_N(\lambda)\}
  =Tr\{\partial_\lambda G_D(\lambda) -
    \partial_\lambda G_N(\lambda)\}\, ,
\end{equation}
we get
 \begin{equation}\label{trazas}
  \begin{array}{c}
    Tr\{\partial_\lambda G_D(\lambda) -
    \partial_\lambda G_N(\lambda)\}
    =\\ \\
    \displaystyle{=-
    \frac{2 g}{\lambda^2}-
    \frac{2 g}{\lambda}\left[
    \frac{J_{g+\frac 1 2  }(\lambda)}{J_{g-\frac 1 2}(\lambda)}+
    \frac{J_{-g - \frac 1 2}(\lambda)}{J_{-g+\frac 1 2}(\lambda)}
    \right] + \left[
    \frac{J_{g+\frac 1 2  }^2(\lambda)}{J_{g-\frac 1
    2}^2(\lambda)}-
    \frac{J_{-g-\frac 1 2}^2(\lambda)}{J_{-g+\frac 1 2}^2(\lambda)}
    \right]=
    } \\ \\
    \displaystyle{=
    \frac{2\,g}{{\lambda }^2} +
  {\left( \frac{1}{2\,\lambda } +
      \frac{{J'_{\frac{1}{2} - g}}(
         \lambda )}{J_{\frac{1}{2} - g}(\lambda )}
      \right) }^2 - {\left( \frac{1}
       {2\,\lambda } +
      \frac{{J'_{g- \frac{1}{2}}}( \lambda )}
      {J_{ g- \frac{1}{2}}( \lambda )} \right) }^2}\, .
  \end{array}
 \end{equation}

Finally, we can also write
\begin{equation}\label{TrG2}
  Tr\{G^2(\lambda)\} = Tr\{\partial_\lambda G_D(\lambda)\} -
  \partial_\lambda \Big[ \tau(\lambda)\,
  Tr\{G_D(\lambda)-G_N(\lambda)\}\Big]\, .
\end{equation}

\bigskip

\section{Asymptotic expansion for the trace of the
resolvent}\label{Asymptotic-expansion}

Using the Hankel asymptotic expansion for Bessel functions
\cite{A-S} (see Appendix \ref{Hankel}), we get for the first term
in the right hand side of \linebreak Eq.\ (\ref{TrG2})
\begin{equation}\label{asymp-trGD-upp}\begin{array}{c}
    \displaystyle{Tr\{\partial_\lambda G_D(\lambda)\} \sim
  \sum_{k=2}^\infty \frac{A_k(g,\sigma)}{\lambda^k}
  = }\\ \\ =\displaystyle{
   - \frac{g }
   {{\lambda }^2} + i\,\sigma\,
  \frac{g\,\left(  g -1 \right)}{{\lambda }^3} -
  \frac{3}{2}\, \frac{
      g\,\left( g  -1 \right)}{{\lambda }^4} \, +}
      \\ \\ \displaystyle{  + i \, \sigma\,
  \frac{\left( g -3  \right) \,
     \left(  g -1 \right) \,g\,
     \left(  g+  2 \right) }{2\,{\lambda }^5} +
   {{O}\left(\frac{1}
      {\lambda }\right)}^6
    }\, ,
\end{array}
\end{equation}
where $\sigma = 1$ for $\Im(\lambda)>0$, and $\sigma = -1$ for
$\Im(\lambda)<0$. The coefficients in this series can be
straightforwardly evaluated from Eqs.\ (\ref{P+Q}) and (\ref{T}).
Notice that $A_k(g,-1)=A_k(g,1)^*$, since $A_{2k}(g,1)$ is real
and $A_{2k+1}(g,1)$ is pure imaginary.

Similarly, from Eqs.\ (\ref{TrGD-GN}), (\ref{trazas}) and
(\ref{JprimasobreJasymp}) we  simply get
\begin{equation}\label{trdifasymp}
   Tr\{G_D(\lambda) -
    G_N(\lambda)\} \sim - \frac{2 g}{\lambda}
\end{equation}
and
\begin{equation}\label{trdifderasymp}
   Tr\{\partial_\lambda G_D(\lambda) -
    \partial_\lambda G_N(\lambda)\} \sim
    \frac{2 g}{\lambda^2}\, .
\end{equation}

\bigskip

On the other hand, taking into account Eq.\ (\ref{JsobreJup}), we
have
\begin{equation}\label{tau-asymp}\begin{array}{c}
   \tau(\lambda) \sim \displaystyle{1 - \frac{1}
   {1 - \displaystyle{\frac{e^
         {\sigma \,i\,
           \pi \, \left( \frac{1}{2} - g \right) }
           \,{\lambda }^{2\,g}}{\rho(\alpha,\beta)
        }} }
        \sim }\\ \\  \sim   \left\{
  \begin{array}{l}
  \displaystyle{
    -\sum_{k=1}^\infty \left( \frac{e^{\sigma\, i\, \pi\, (\frac 1 2 -g)}
    \lambda^{2 g}}{\rho(\alpha,\beta)}\right)^k , \ {\rm for}\
    -\frac 1 2 < g < 0\, ,} \\ \\
      \displaystyle{
    \sum_{k=0}^\infty \left(\rho(\alpha,\beta)\,
    {e^{-\sigma\, i\, \pi\, (\frac 1 2 -g)} \,
    \lambda^{-2 g}}\right)^k , \ {\rm for}\
    0 < g < \frac 1 2\, ,}
  \end{array}
  \right.
\end{array}
\end{equation}
where $\sigma =1$ ($\sigma =-1$) corresponds to $\Im(\lambda)>0$
($\Im(\lambda)<0$). Notice the appearance of non integer,
$g$-dependent, powers of $\lambda$ in this asymptotic expansion.

Similarly
\begin{equation}\label{tauprima}\begin{array}{c}
    \tau'(\lambda) \sim \displaystyle{-
  {\left( 1 - \frac{e^
         {\sigma\, i\, \pi   \,\left( \frac{1}{2} -
         g \right) }\,{\lambda }^{2\,g}}{\rho(\alpha,\beta)
        } \right) }^{-2}\,
        \frac{e^
     {\sigma\, i\, \pi  \,\left( \frac{1}{2} -
     g \right) }\,2\,g\,{\lambda }^{ 2\,g-1 }}
    {\rho(\alpha,\beta) }
    \sim } \\ \\
   \sim \displaystyle{
   \left\{\begin{array}{l}
     \displaystyle{
    - \frac{2\,g}{\lambda}\,\sum_{k=1}^\infty k
    \left( \frac{e^{\sigma\, i\, \pi\, (\frac 1 2 -g)}\,
    \lambda^{2 g}}{\rho(\alpha,\beta)}\right)^k , \ {\rm for}\
    -\frac 1 2 < g < 0\, ,} \\ \\
           \displaystyle{
    -\frac{2\,g}{\lambda}\,
    \sum_{k=1}^\infty k  \left(\rho(\alpha,\beta)\,
    {e^{-\sigma\, i\, \pi\, (\frac 1 2 -g)} \,
    \lambda^{-2 g}}\right)^k , \ {\rm for}\
    0 < g < \frac 1 2\,  ,}
   \end{array}
   \right. }
\end{array}
\end{equation}
which are the term by term derivatives of the corresponding
asymptotic series in Eq.\ (\ref{tau-asymp}).

Therefore, we have
\begin{equation}\label{des-asymp-deriv-tau-Tr}\begin{array}{c}
   \partial_\lambda \Big[ \tau(\lambda)\,
  Tr\{G_D(\lambda)-G_N(\lambda)\}\Big]
  \sim \\ \\
   \sim \left\{
  \begin{array}{l}
  \displaystyle{
    2\, g \sum_{k=1}^\infty
    \left( \frac{e^{\sigma\, i\, \pi\, (\frac 1 2 -g)}
    }{\rho(\alpha,\beta)}\right)^k
    \left(2\, g \, k -1\right) \lambda^{2\, g \, k -2}, \ {\rm for}\
    -\frac 1 2 < g < 0\, ,} \\ \\
      \displaystyle{ 2\, g
    \sum_{k=0}^\infty \left(\rho(\alpha,\beta)\,
    {e^{-\sigma\, i\, \pi\, (\frac 1 2 -g)}
    }\right)^k
    \left(2\, g \, k + 1\right) \lambda^{-2\, g \, k -2}
    , \ {\rm for}\
    0 < g < \frac 1 2\, .}
  \end{array}
  \right.
\end{array}
\end{equation}
Notice the $g$-dependent powers of $\lambda$ appearing in these
asymptotic expansions.

\bigskip

\section{The $\zeta$ and $\eta$ functions} \label{spectral-functions}

The $\zeta$-function for a general self-adjoint extension of $D_x$
is defined, for $\Re(s)>1$, as \cite{Seeley}
\begin{equation}\label{zeta}
  \zeta(s)=- \frac{1}{2\,\pi\,i} \oint_{\mathcal{C}}
  \frac{\lambda^{1-s}}{s-1} \, Tr\left\{G^2(\lambda)\right\}
  \, d\lambda\, ,
\end{equation}
where the curve $\mathcal{C}$ encircles counterclockwise the
spectrum of the operator, keeping to the left of the origin.
According to Eq.\ (\ref{TrG2}), we have
\begin{equation}\label{zeta1}
  \zeta(s)= \zeta^D(s)+\frac{1}{2\,\pi\,i} \oint_{\mathcal{C}}
  \frac{\lambda^{1-s}}{s-1} \, \partial_\lambda \Big[ \tau(\lambda)\,
  Tr\{G_D(\lambda)-G_N(\lambda)\}\Big]
  \, d\lambda\, ,
\end{equation}
where $\zeta^D(s)$ is the $\zeta$-function for the $D$-extension.

Since the negative eigenvalues of the self-adjoint extension of
$D_x$ characterized by the pair $(\alpha,\beta)$,
$D_x^{(\alpha,\beta)}$, are minus the positive eigenvalues
corresponding to the extension $D_x^{(\alpha,-\beta)}$ (as
discussed in the Section \ref{the-spectrum}), we define a partial
$\zeta$-function through a path of integration encircling the
positive eigenvalues only,
\begin{equation}\label{zeta+}\begin{array}{c}
  \displaystyle{  \zeta_+^{(\alpha,\beta)}(s)=
  \frac{1}{2\,\pi\,i} \int_{-i\,\infty+0}^{i\,\infty+0}
  \frac{\lambda^{1-s}}{s-1} \, Tr\left\{G^2(\lambda)\right\}
  \, d\lambda =
    } \\ \\
    \displaystyle{
    = \zeta_+^{D}(s)
    -\frac{1}{2\,\pi\,i} \int_{-i\,\infty+0}^{i\,\infty+0}
  \frac{\lambda^{1-s}}{s-1} \,
  \partial_\lambda \Big[ \tau(\lambda)\,
  Tr\{G_D(\lambda)-G_N(\lambda)\}\Big]
  \, d\lambda
    } \, ,
\end{array}
\end{equation}
where $\zeta_+^{D}(s)$ is the partial $\zeta$-function for the
$D$-extension.

We can also write
\begin{equation}\label{zeta+1}\begin{array}{c}
  \displaystyle{\zeta_+^{(\alpha,\beta)}(s)=
    \frac{1}{2\,\pi} \int_{1}^{\infty}
    e^{i\,\frac{\pi}{2}\,(1-s)}\,
  \frac{\mu^{1-s}}{s-1} \,
  Tr\left\{G^2(e^{i\,\frac{\pi}{2}}\, \mu)\right\}
  \, d\mu\,  +
  } \\ \\
  \displaystyle{
    +\frac{1}{2\,\pi} \int_{1}^{\infty}
    e^{- i\,\frac{\pi}{2}\,(1-s)}\,
  \frac{\mu^{1-s}}{s-1} \,
  Tr\left\{G^2(e^{-i\,\frac{\pi}{2}}\, \mu)\right\}
  \, d\mu + \frac{h_1(s)}{s-1}\, ,
  }
\end{array}
\end{equation}
where $h_1(s)$ is an entire function. Therefore, in order to
determine the poles of $\zeta_+^{(\alpha,\beta)}(s)$, we can
subtract and add a partial sum of the asymptotic expansion
obtained in the previous Section to
$Tr\left\{G^2(\lambda)\right\}$ in the integrands in the right
hand side of Eq.\ (\ref{zeta+1}).

\bigskip

In so doing, we get for the $D$-extension and for a real $s>1$
\begin{equation}\label{zetaD+1}\begin{array}{c}
  \displaystyle{ \zeta_+^{D}(s) =
    \frac{1}{2\,\pi\,(s-1)} \int_{1}^{\infty}
    e^{i\,\frac{\pi}{2}\,(1-s)}\,
  {\mu^{1-s}} \,
  \left\{ \sum_{k=2}^{N} e^{-i\,\frac{\pi}{2}\,k}\,
  A_k(g,1) \, \mu^{-k} \right\}
  \, d\mu\,  +
  } \\ \\
  \displaystyle{
    \frac{1}{2\,\pi\,(s-1)} \int_{1}^{\infty}
    e^{-i\,\frac{\pi}{2}\,(1-s)}\,
  {\mu^{1-s}} \,
  \left\{ \sum_{k=2}^{N} e^{i\,\frac{\pi}{2}\,k}\,
  A_k(g,1)^* \, \mu^{-k} \right\}
  \, d\mu + \frac{h_2(s)}{s-1}=
  } \\ \\
  \displaystyle{
  = \frac{1}{\pi\,(s-1)}\,\sum_{k=2}^{N}\frac{1}{s-(2-k)}\, \Re \left\{
  e^{i\,\frac{\pi}{2}\,(1-s-k)}\,A_k(g,1) \right\}
  + \frac{h_2(s)}{s-1}\, ,
  }
\end{array}
\end{equation}
where $h_2(s)$ is an analytic function in the open half plane
$\Re(s)> 1-N$.

Consequently, the meromorphic extension of $\zeta_+^{D}(s)$
presents a simple pole at $s=1$ (see Eq.\ (\ref{zeta+})), with a
residue given by (see Eq.\ (\ref{traza-derivGD}))
\begin{equation}\label{residuos=1}
  \left. {\rm Res}\,\zeta_+^{D}(s) \right|_{s=1}
  =  \frac{1}{2\, \pi \, i} \int_{-i\, \infty+0}^{i\, \infty+0}
  \lambda^0 \, \partial_\lambda \left\{
  \frac{J_{g+\frac 1 2}(\lambda)}{J_{g-\frac 1 2}(\lambda)}
  \right\}\, d\lambda = \frac 1 \pi\, ,
\end{equation}
where we have used Eqs.\ (\ref{JsobreJup1}) and
(\ref{asymp-cociente}).

It also presents simple poles at $s=2-k$, for $k=2,3,\dots$, with
residues given by
\begin{equation}\label{otros-residuos}
  \left. {\rm Res}\,\zeta_+^{D}(s) \right|_{s=2-k}
  = \frac{\Re\left\{i\, A_k(g,1)\right\}}{(k-1)\, \pi}  \, ,
\end{equation}
with the coefficients $A_k(g,1)$ given in Eq.\
(\ref{asymp-trGD-upp}). In particular, notice that these residues
vanish for even $k$.

\bigskip

For a general self-adjoint extension $D_x^{(\alpha,\beta)}$, we
must also consider  the singularities coming from the asymptotic
expansion of $\partial_\lambda [ \tau(\lambda)$
$Tr\{G_D(\lambda)-G_N(\lambda)\}]$ in Eq.\
(\ref{des-asymp-deriv-tau-Tr}). For definiteness, let us consider
in the following the case $-\frac 1 2 < g <0$ (the case $0<g<\frac
1  2$ leads to similar results).

\bigskip

From Eq.\ (\ref{zeta+}), and taking into account Eq.\
(\ref{zeta+1}), for real $s>1$ we can write
\begin{equation}\label{zetadif}\begin{array}{c}
    \zeta_+^{(\alpha,\beta)}(s)-\zeta_+^{D}(s)=
    \displaystyle{\frac{h_3(s)}{s-1}} \, -\\ \\
  \displaystyle{-
     \frac{ g}{\pi\,(s-1)} \int_{1}^{\infty}
    e^{i\,\frac{\pi}{2}\,(-s-1)}\,
  {\mu^{1-s}} \,
  \left\{ \sum_{k=1}^{N}\left(\frac{ e^{i\,\frac{\pi}{2}}}
  {\rho(\alpha,\beta)}\right)^k\,
  (2\,g\,k-1) \, \mu^{2\,g\,k-2} \right\}
  \, d\mu
  } \\ \\
   -\displaystyle{
       \frac{g}{\pi\,(s-1)} \int_{1}^{\infty}
    e^{-i\,\frac{\pi}{2}\,(-s-1)}\,
  {\mu^{1-s}} \,
  \left\{ \sum_{k=1}^{N}\left(\frac{ e^{-i\,\frac{\pi}{2}}}
  {\rho(\alpha,\beta)}\right)^k\,
  (2\,g\,k-1) \, \mu^{2\,g\,k-2} \right\}
  \, d\mu
  } \\ \\
  \displaystyle{
  = \frac{-\,2\,g}{\pi\,(s-1)}\, \sum_{k=1}^{N}\,
  \left(\frac{2\,g\,k-1}{s-2\,g\,k}\right)\, \Re \left\{
  \frac{e^{i\,\frac{\pi}{2} (k-s-1)}}{\rho(\alpha,\beta)^k}\right\}
  + \displaystyle{\frac{h_3(s)}{s-1}}\, ,
  }
\end{array}
\end{equation}
where $h_3(s)$ is holomorphic for $\Re(s)>2\,g\,(N+1)$.

Therefore,
$\left(\zeta_+^{(\alpha,\beta)}(s)-\zeta_+^{D}(s)\right)$ has a
meromorphic extension which presents a simple pole at $s=1$, with
a vanishing residue,
\begin{equation}\label{res-dif-s=1}\begin{array}{c}
    \left. {\rm Res}\,\left(\zeta_+^{(\alpha,\beta)}(s)-
  \zeta_+^{D}(s)\right) \right|_{s=1}
  = \\ \\ \displaystyle{
  = -\frac{1}{2\,\pi\,i} \int_{-i\,\infty+0}^{i\,\infty+0}
  {\lambda^{0}}\,
  \partial_\lambda \Big[ \tau(\lambda)\,
  Tr\{G_D(\lambda)-G_N(\lambda)\}\Big]
  \, d\lambda =0}\, ,
\end{array}
\end{equation}
as follows from Eqs.\ (\ref{trdifasymp}) and (\ref{tau-asymp}).

Notice also the presence of simple poles located at negative non
integer $g$-dependent positions, $s=2\,g\,k=- 2\,|g|\,k$, for
$k=1,2,\dots$, with residues which also depend on the self-adjoint
extension, given by
\begin{equation}\label{res-g-dep}\begin{array}{c}
      \left. {\rm Res}\,\left\{\zeta_+^{(\alpha,\beta)}(s)-
  \zeta_+^{D}(s)\right\} \right|_{s=2\,g\,k} =
    \\ \\
    = \displaystyle{ \frac{-\, 2\,g}{\pi\,\rho(\alpha,\beta)^k}
    \ {\sin\left[\left(\frac{1}{2} -g\right)k\,\pi\right]}
  }\, .
\end{array}
\end{equation}

\bigskip

Now, taking into account our comment after Eq.\
(\ref{eigenvalues-neg}), we get for the complete $\zeta$-function
\begin{equation}\label{zeta-completa}
  \zeta^{(\alpha,\beta)}(s) = \zeta_+^{(\alpha,\beta)}(s)+
  e^{-i\,\pi\,s} \, \zeta_+^{(\alpha,-\beta)}(s)\, .
\end{equation}
{  In particular, for the $\alpha=0$ extension we get
\begin{equation}\label{zeta-alpha0}
  \zeta^D(s) = \left(1 + e^{-i\,\pi\,s}\right)
  \zeta_+^{D}(s),
\end{equation}
since the spectrum of $D_x^{(0,1)}$ is symmetric with respect to
the origin (see Eq.\ (\ref{alpha0})). Then one concludes that
$\zeta^D(s)$ has vanishing residues. Indeed,  from Eq.\
(\ref{otros-residuos}), the residue at $s=2-k$ vanishes for $k$
even, and for $k=2\,l+1$, with $l=0,1,2,\dots$, we have
\begin{equation}\label{residuos-nulos}
   \left. {\rm Res}\,\left\{\zeta^{D}(s)\right\}
   \right|_{s=1-2\,l} =
   \left(1 + e^{-i\,\pi\left(1-2\,l\right)}\right)
   \left. {\rm Res}\,\left\{\zeta_+^{D}(s)\right\}
   \right|_{s=1-2\,l} =0\, .
\end{equation}
}

On the other hand,  for a general self-adjoint extension, the
singularities of $\zeta^{(\alpha,\beta)}(s)$ are simple poles
located at $s=2\,g\,k<0$, for $k=1,2,\dots$, with residues
\begin{equation}\label{residues-zeta-completa}\begin{array}{c}
 \displaystyle{
  \left. {\rm Res}\,\left\{\zeta^{(\alpha,\beta)}(s)-
  \zeta^{D}(s)\right\} \right|_{s=2\,g\,k} =
  }\\ \\
  = \displaystyle{\left. {\rm Res}\,\left\{
  \left[\zeta_+^{(\alpha,\beta)}(s)-
  \zeta_+^{D}(s)\right] + e^{-i\,\pi\,s} \,
  \left[\zeta_+^{(\alpha,-\beta)}(s)-
  \zeta_+^{D}(s)\right]
  \right\} \right|_{s=2\,g\,k} } =\\ \\
  \displaystyle{=
  (-1)^k \,
  \frac{2\,g}{\pi} \
  \frac{\sin(2\,g\,k\,\pi)}{\rho(\alpha,\beta)^k}}\,
  e^{i\,\pi\left(\frac 1 2 -g\right) k}\, ,
\end{array}
\end{equation}
where we have used $\rho(\alpha,-\beta)=-\rho(\alpha,\beta)$, from
Eq.\ (\ref{rho}).

\bigskip

Similarly, for the spectral asymmetry \cite{APS} we have
\begin{equation}\label{eta-completa}
  \eta^{(\alpha,\beta)}(s) = \zeta_+^{(\alpha,\beta)}(s)-
  \zeta_+^{(\alpha,-\beta)}(s)\, .
\end{equation}
{ In particular, $\eta^{(0,1)}(s)\equiv 0 \equiv \eta^{(1,0)}(s)$,
since these spectra are symmetric (see Eqs.\ (\ref{alpha0}) and
(\ref{eigen-beta0})). }

For a general self-adjoint extension and $-\frac 1 2 < g <0$,
$\eta^{(\alpha,\beta)}(s)$ presents simple poles at $s=2\,g\,k$,
for $k=1,2,\dots$, with residues given by
\begin{equation}\label{residues-eta}
\displaystyle{
  \left. {\rm Res}\,\left\{\eta^{(\alpha,\beta)}(s)\right\}
  \right|_{s=2\,g\,k} = \left[(-1)^k-1\right]
  \frac{2\,g}{\pi} \,
  \frac{\sin\left[\left(\frac{1}{2}-g\right)
  \,k\,\pi\right]}{\rho(\alpha,\beta)^k}}\, ,
\end{equation}
which vanish for even $k$.

\bigskip

For the case $0<g<\frac 1 2$, an entirely similar calculation
shows that
$\left(\zeta_+^{(\alpha,\beta)}(s)-\zeta_+^{D}(s)\right)$ has a
meromorphic extension which presents simple poles at negative non
integer $g$-dependent positions, $s=-2\,g\,k$, for $k=1,2,\dots$,
with residues depending on the self-adjoint extension, given by
\begin{equation}\label{res-gpos-dep}\begin{array}{c}
      \left. {\rm Res}\,\left\{\zeta_+^{(\alpha,\beta)}(s)-
  \zeta_+^{D}(s)\right\} \right|_{s=-2\,g\,k} =
    \\ \\
    = \displaystyle{ -\,\frac{2\,g}{\pi}\,\rho(\alpha,\beta)^k
    \ {\sin\left[\left(\frac{1}{2} -g\right)k\,\pi\right]}
  }\, .
\end{array}
\end{equation}
From this result, it is immediate to get the residues for the
$\eta$ and $\zeta$-functions. One gets the same expressions as in
the right hand sides of Eqs.\ (\ref{residues-zeta-completa}) and
(\ref{residues-eta}), with $\rho(\alpha,\beta)$ and
$e^{i\,\pi\left(\frac 1 2 -g\right) k}$ replaced by their
inverses.

\bigskip

{ Let us remark that when neither $\alpha$ nor $\beta$ is 0, the
residue of $\zeta_+^{(\alpha,\beta)}$ at $s=-2\,|g|\,k$ is a
constant times $(\beta/\alpha)^{k\,{\rm sign}(g)}$. This is
consistent with the behavior of $D_x$ under the scaling isometry
$Tu(x)=c^{1/2}\,u(c\,x)$ taking $\mathbf{L_2}(0,1)\rightarrow
\mathbf{L_2}(0,1/c)$. The extension $D_x^{(\alpha,\beta)}$ is
unitarily equivalent to the operator
$(1/c)\dot{D}_x^{(\alpha',\beta')}$ similarly defined on
$\mathbf{L_2}(0,1/c)$, with $\alpha' = c^{-g}\, \alpha$ and
$\beta' = c^g \, \beta$:
\begin{equation}\label{isometry}
  T\,D_x^{(\alpha,\beta)} = \frac 1 c \,
  \dot{D}_x^{(\alpha',\beta')}\, T\, .
\end{equation}
Notice that only for the extensions with $\alpha=0$ or $\beta=0$
the boundary condition at the singular point $x=0$, Eq.\
(\ref{BC2}), is left invariant by this scaling.

Therefore, we have for the partial $\zeta$-function of the scaled
problem
\begin{equation}\label{zetas-isometry}
  \dot{\zeta}_+^{(\alpha',\beta')}(s)=
   c^{-s}\,\zeta_+^{(\alpha,\beta)}(s)\, ,
\end{equation}
and for the residues
\begin{equation}\label{zetas-isometry-residues}
   \left. {\rm Res}\,\left\{\dot{\zeta}_+^{(\alpha',\beta')}(s)
   \right\} \right|_{s=-2\,|g|\,k} = c^{2\,|g|\,k}
   \left. {\rm Res}\,\left\{{\zeta}_+^{(\alpha,\beta)}(s)
   \right\} \right|_{s=-2\,|g|\,k}\, .
\end{equation}
The factor $c^{2\,|g|\,k}$ exactly cancels the effect the change
in the boundary condition at the singularity has on
$\rho(\alpha,\beta)$,
\begin{equation}\label{change-in-rho}
  \rho(\alpha,\beta)^{k\,{\rm sign}(g)}
  =c^{-2\,|g|\,k}\, \rho(\alpha',\beta')^{k\,{\rm sign}(g)}\, .
\end{equation}
Thus the difference between the intervals $(0,1)$ and $(0,1/c)$
has no effect on the structure of these residues, which presumably
are determined locally in a neighborhood of $x=0$. }

\bigskip

Finally, let us point out that these anomalous poles are not
present in the $g=0$ case. Indeed, in this case $\tau(\lambda)$ in
Eq.\ (\ref{taudelambda}) has a constant asymptotic expansion,
while $Tr\{G_D(\lambda)-G_N(\lambda)\} \sim 0$ (see Eq.\
(\ref{trdifasymp})). Moreover, the residues of the poles coming
from $\zeta_+^{D}(s)$ are all zero (see Eqs.\
(\ref{otros-residuos}) and (\ref{asymp-trGD-upp})), except for the
one at $s=1$, with residue $1/\pi$ (see Eq.\ (\ref{residuos=1})).

\bigskip

Consequently, the presence of poles in the spectral functions
located at non integer positions is a consequence of the singular
behavior of the 0-th order term in $D_x$ near the origin, together
with a boundary condition which is not invariant under scaling.

\bigskip

\section{Comments on the second order case}\label{second-order}

In this Section we briefly describe similar results one can obtain
for the self-adjoint extensions of the second order differential
operator
\begin{equation}\label{Delta}
  \Delta_x = - \partial_x^2 + \frac{g(g-1)}{x^2}\, ,
\end{equation}
with $- \frac 1 2 < g < \frac 1 2$, defined on a set of functions
satisfying $\phi(1)=0$ and behaving as
\begin{equation}\label{BC-second}
  \phi(x)= C_1 \, x^g + C_2\, x^{1-g} + O(x^{3/2}) \, ,
\end{equation}
where the coefficients $C_{1,2}$ are constrained as in Eq.\
(\ref{BC2}).

It can be shown that the spectrum of the self-adjoint extension
$\Delta_x^{(\alpha,\beta)}$ is determined by a relation similar to
Eq.\ (\ref{eigenvalues-pos}):
\begin{equation}\label{spectra-second}
  \mathcal{F}(\mu):= \frac 1 \mu\, F(\mu) = \varrho (\alpha,\beta)\, ,
\end{equation}
where the constant
\begin{equation}\label{varrho}
  \varrho(\alpha,\beta) := \left(\frac \beta \alpha \right)
  2^{2\,g-1} \, \frac{\Gamma(\frac 1 2 +g)}
  {\Gamma(\frac 3 2 -g)} \, .
\end{equation}

Also in this case,  $\alpha = 0$ and $\beta =0$ correspond to two
scale invariant boundary conditions at the singularity. For these
two limiting extensions, it is easily seen from Eqs.\
(\ref{ec-dif-g-diag}), (\ref{near0-D}) and (\ref{near0-N}) that
the entry $G_{11}(x,y;\mu)$ in the resolvent of our first order
operator $D_x^{(\alpha,\beta)}$ is $\mu$ times the corresponding
resolvent of $\Delta_x^{(\alpha,\beta)}$ at $\lambda = \mu^2$,
\begin{equation}\label{resolv-second}
  \mathcal{G}_{D,N}(x,y;\mu^2) =
  \frac 1 \mu \, G_{11}^{D,N}(x,y;\mu)\, .
\end{equation}
The resolvent for a general self-adjoint extension
$\Delta_x^{(\alpha,\beta)}$ is constructed as a convex linear
combination of $\mathcal{G}_{D}(\mu^2)$ and
$\mathcal{G}_{N}(\mu^2)$ as in (\ref{linear-comb}), with a
coefficient
\begin{equation}\label{tau-second}
  \tau(\mu) =  \displaystyle{
  \frac{1}{1-
  \displaystyle{
  \frac{\varrho(\alpha,\beta)}{\mathcal{F}(\mu)}} } }\, .
\end{equation}

Following the methods employed for the first order case, one can
show that the $\zeta$-function associated to
$\Delta_x^{(\alpha,\beta)}$ also displays anomalous poles located
at $s=-\left(\frac 1 2 -g \right)k$, with $k=1,2,\dots$, which
implies the presence of anomalous powers $t^{\left(\frac 1 2 -g
\right)k}$ in the heat trace small-$t$ asymptotic expansion. The
residues at these poles, and the corresponding heat trace
coefficients are similarly evaluated. More details on this
calculation will be reported elsewhere.

\vskip 1cm

\noindent {\bf Note added in proof:} It has come belatedly to our
attention that the article by Edith A. Mooers, ``Heat kernel
asymptotics on manifolds with conic singularities", J. Anal. Math.
vol 78 (1999) 1-36, gives the first ``unusual" term in the
expansion of the Laplacian on the half line with a domain which is
not scaling invariant. That article also gives a construction
which in principle would give the complete expansion in the case
of a manifold with isolated conic singularities, for an arbitrary
self-adjoint realization of the Laplacian. For the case considered
here, by contrast, the present results are simpler and more
complete, as they treat the first order case and the eta
invariant, and give more explicit coefficients.

\bigskip

\section*{Acknowledgements}

HF and PAGP acknowledge support from Universidad Nacional de La
Plata (grant 11/X298) and CO\-NI\-CET (grant 0459/98), Argentina.
They also acknowledge support from CLAF-ICTP (grant 196/02).

MAM acknowledge support from Universidad Nacional de La Plata
(grant 11/X228), Argentina.

\bigskip

\appendix

\section{Evaluation of the traces}\label{integrals}

In this Appendix we briefly describe the evaluation of the traces
appearing in Section \ref{trace-resolvent}.

We need to compute
\begin{equation}\label{int-Tr-GD-GN}\begin{array}{c}
  Tr\{G_D(\lambda)-G_N(\lambda)\}= \\
  = \displaystyle{ \int_0^1
  \Big[tr\left\{G_D(x,x;\lambda) \right\}
  -tr\left\{G_N(x,x;\lambda) \right\}\Big]
  dx\, . }
\end{array}
\end{equation}

Let us first consider the contribution of $G_N(\lambda)$ to this
integral. From Eq.\ (\ref{GN11}) and (\ref{GN22}) we get for the
matrix trace of $G_N(\lambda)$ on the diagonal
\begin{equation}\label{trGN-matrix}\begin{array}{c}
    tr\left\{G_N(x,x;\lambda) \right\}
  =- \, \displaystyle{\frac{\pi \,x\,\lambda\,\sec (g\,\pi ) }
    {2\,J_{\frac{1}
       {2} - g}(\lambda )}}\,
    \left\{ {J_{- \frac{1}{2}
              - g}(
          x\,\lambda )}^2\,
       J_{-
            \frac{1}{2}   + g}(\lambda ) + \right.
            \\ \\
  \left. +
      J_{\frac{1}
          {2} - g}(x\,\lambda )\,
       \left[ J_{
           \frac{1}{2} - g}(x\,\lambda )\,
          J_{- \frac{1}{2}
             + g}(\lambda )
          - J_{\frac{1}{2} - g}(\lambda )
           \,J_{- \frac{1}{2}
             + g}(x\,\lambda ) \right]  + \right.
             \\ \\ \left. +
      J_{- \frac{1}{2}   - g}(x\,\lambda )\,
       J_{\frac{1}
          {2} - g}(\lambda )\,
       J_{\frac{1}
          {2} + g}(x\,\lambda )
      \right\}\, ,
\end{array}
\end{equation}
an integrable expression behaving as
\begin{equation}\label{TRGN-matrix-en0}\begin{array}{c}
  tr\left\{G_N(x,x;\lambda) \right\}
    = \\ \\ \displaystyle{ =
  {x^{-2\,g}}\left\{{-\left( \frac{4^g\,\pi \,\sec (g\,\pi )\,
         J_{g-\frac{1}{2}}(\lambda )
         }{{\lambda }^{2\,g}\,
         {\Gamma(\frac{1}{2} - g)}^2 \,
         J_{\frac{1}{2} - g}(\lambda )}
       \right)  +{O}(x)}\right\} +  {O}(x)}
\end{array}
\end{equation}
near the origin.

Therefore, it is sufficient to know the primitives
\cite{Mathematica}
\begin{equation}\label{nu-nu}
  \int x\, J_\nu^2(\lambda\,x)\,dx
  =\frac{x^2}{2}\,\left\{ {J_\nu(x\,\lambda )}^2 -
      J_{\nu-1}(x\,\lambda )\,
       J_{\nu+1}(x\,\lambda ) \right\}
\end{equation}
and
\begin{equation}\label{nu-menosnu}\begin{array}{c}
 \displaystyle{ \int x\, J_\nu(\lambda\,x)\,
 J_{-\nu}(\lambda\,x)\,dx= }\\ \\
     =\displaystyle{\frac{ - {\nu }^2}
      {{\lambda }^2\,
      \Gamma(1 - \nu )\,
      \Gamma(1 + \nu )}\,
      \left[ \,_1 F_2
          \left(\{ - {1}/
            {2}  \} ,
         \{ -\nu ,\nu \} ,
         -x^2 \, {\lambda }^2\right)-1  \right]}\, ,
\end{array}
\end{equation}
where
\begin{equation}\label{1F2}\begin{array}{c}
   _1 F_2 \left(\{ - {1}/
            {2}  \} ,
         \{ -\nu ,\nu \} ,
         -x^2 \, {\lambda }^2\right)
  \displaystyle{
  =- \frac{ \pi \,x^2\,
      {\lambda }^2\,
       \,\csc (\pi \,\nu )
       }{4\,\nu }\, \times }\\ \\
      \left\{ J_{-1 - \nu }(x\,\lambda )\,
         J_{-1 + \nu }(x\,\lambda ) +
        2\,J_{-\nu }(x\,\lambda )\,
         J_{\nu }(x\,\lambda ) +
        J_{1 - \nu}(x\,\lambda )\,
         J_{1 + \nu }(x\,\lambda )
        \right\}\, .
\end{array}
\end{equation}

These primitives, together with the relation
\begin{equation}\label{prop-J-Bessel}
  J_{\nu-1}(z)+J_{\nu+1}(z)
  =\frac{2 \nu}{z} J_{\nu}(z)\, ,
\end{equation}
necessary to simplify the intermediate results, eventually lead to
\begin{equation}\label{TRGN}
  I_N(\lambda):=\displaystyle{ \int_0^1
  tr\left\{G_N(x,x;\lambda) \right\}\,
  dx
  =-\,\frac{2\,g}{\lambda } -
  \frac{J_{- g-\frac{1}{2}  }(\lambda )}{J_{- g+\frac{1}{2} }(\lambda
  )}\, . }
\end{equation}

\bigskip

Similarly, for the matrix trace of $G_D(\lambda)$ on the diagonal
we have
\begin{equation}\label{trGD}\begin{array}{c}
     tr\left\{G_D(x,x;\lambda) \right\}= \\ \\
     =     \displaystyle{\frac{\pi \,x\,\lambda \,\sec (g\,\pi )}
    {2\,J_{-\frac{1}{2} + g}(\lambda )}}
    \,
    \left\{- J_{\frac{1}{2} - g}(x\,\lambda )
        \,J_{- \frac{1}{2}
             + g}(\lambda )\,
       J_{-  \frac{1}{2}   + g}(x\,\lambda )\, + \right.
       \\ \\
       +
      J_{-\frac{1}{2}  - g}(x\,\lambda )\,
       J_{- \frac{1}{2}   + g}(\lambda )\,
       J_{\frac{1}
          {2} + g}(x\,\lambda ) \, +
       \\ \\ \left.  +
      J_{\frac{1}
          {2} - g}(\lambda )\,
       \left( {J_{-\frac{1}{2}+ g}(
            x\,\lambda )}^2 +
         {J_{\frac{1}{2} + g}(
            x\,\lambda )}^2 \right)
      \right\}\, ,
\end{array}
\end{equation}
which behaves as
\begin{equation}\label{trGD-en0}\begin{array}{c}
  tr\left\{G_D(x,x;\lambda) \right\}
  = \\ \\
  =\displaystyle{x^{2\,g}\,\left\{ \frac{\pi \,{\lambda }^{2\,g}
    \,\sec (g\,\pi )\,
       J_{\frac{1}{2} - g}(\lambda )}{4^g\,
       {\Gamma(\frac{1}{2} + g)}^2 \,
       J_{g-\frac{1}{2}}(\lambda )} +
    O(x) \right\} + O(x)
  \, .}
\end{array}
\end{equation}

The same argument as before leads to
\begin{equation}\label{TRGD}
  I_D(\lambda):=\displaystyle{ \int_0^1
  tr\left\{G_D(x,x;\lambda) \right\}\,
  dx   =
  \frac{J_{ g+\frac{1}{2} }(\lambda )}
      {J_{g- \frac{1}{2} }(\lambda )}\, . }
\end{equation}

\bigskip

Therefore, we get
\begin{equation}\label{Tr-GD-GN-final}
  Tr\{G_D(\lambda)-G_N(\lambda)\}= I_D(\lambda)-I_N(\lambda)\, ,
\end{equation}
as in Eq.\ (\ref{TrGD-GN}).

\bigskip

On the other hand, we have
\begin{equation}\label{ddd}\begin{array}{c}
  \partial_\lambda
  tr\left\{G_D(x,x;\lambda) \right\}=  {O}(x)\, +\\ \\
  \displaystyle{
  x^{2\,g}\,\left\{ \frac{2^{1 - 2\,g}\,
       {\lambda }^{-1 + 2\,g}\,
       \left[ 1 + g\,\pi \,
          J_{\frac{1}{2} - g}(\lambda )\,
          J_{g-\frac{1}{2}}(\lambda )\,
          \sec (g\,\pi ) \right] }{{
          J_{g-\frac{1}{2}}(\lambda )}^2\,
       {\Gamma(\frac{1}{2} + g)}^2} +
    {O}(x) \right\} } \, .
\end{array}
\end{equation}
Then,
\begin{equation}\label{Int-Tr-deriv-GD}
  Tr\{\partial_\lambda G_D(\lambda)\} =
  \displaystyle{ \int_0^1
  \partial_\lambda
  tr\left\{G_D(x,x;\lambda) \right\}\,
  dx   = \partial_\lambda I_D(\lambda)\, , }
\end{equation}
in agreement with Eq.\ (\ref{traza-derivGD}).

\bigskip

\section{The Hankel expansion}\label{Hankel}

In order to develop an asymptotic expansion for the trace of the
resolvent, we  use  the Hankel asymptotic expansion for the Bessel
functions: For $|z|\rightarrow \infty$, with $\nu$ fixed and
$|\arg z|<\pi$, we have \cite{A-S}
\begin{equation}\label{hankel}
  J_\nu(z)\sim \left(\frac{2}{\pi\,z}\right)^{\frac 1 2}
  \left\{ P(\nu,z) \cos \chi(\nu,z)
  -Q(\nu,z) \sin \chi(\nu,z) \right\}\, ,
\end{equation}
where
\begin{equation}\label{chi}
  \chi(\nu,z) = z- \left(\frac \nu 2 + \frac 1 4\right) \pi,
\end{equation}
\begin{equation}\label{P}
  P(\nu,z) \sim \sum_{k=0}^\infty
  \frac{(-1)^k\,\Gamma\left(
  \frac 1 2 + \nu + 2 k\right)}{(2k)! \, \Gamma\left(
  \frac 1 2 + \nu - 2 k\right)} \,
  \frac{1}{\left(2 z\right)^{2 k}}\, ,
\end{equation}
and
\begin{equation}\label{Q}
  Q(\nu,z) \sim \sum_{k=0}^\infty
  \frac{(-1)^k \,\Gamma\left(\frac 1 2+
   \nu + 2 k+1 \right)}{(2 k+1)! \, \Gamma\left(
   \frac 1 2+ \nu - 2 k-1 \right)} \,
    \frac{1}{\left(2 z\right)^{2 k+1}}\, .
\end{equation}

Moreover, $P(-\nu,z)=P(\nu,z)$ and $Q(-\nu,z)=Q(\nu,z)$, since
these functions depend only on $\nu^2$ (see Ref.\ \cite{A-S}, page
364).

Therefore, for $z$ in the upper open  half plane,
\begin{equation}\label{Jupper}
  J_\nu(z)\sim
  \frac{e^{-i z}\, e^{i \pi \left(
  \frac \nu 2 + \frac 1 4 \right)}}{\sqrt{2\pi z}}
  \left\{ P(\nu,z)
  - i \,Q(\nu,z) \right\}\, ,
\end{equation}
while for $z$ in the lower open  half plane we get
\begin{equation}\label{Jlower}
  J_\nu(z)\sim
  \frac{e^{i z}\, e^{-i \pi \left(
  \frac \nu 2 + \frac 1 4 \right)}}{\sqrt{2\pi z}}
  \left\{ P(\nu,z)
  + i \,Q(\nu,z) \right\}\, .
\end{equation}
In these equations,
\begin{equation}\label{P+Q}
  P(\nu,z) \pm i \,Q(\nu,z)\sim
  \sum_{k=0}^\infty \langle \nu , k\rangle
  \, \left(\frac{\pm i}{2 z}\right)^k\, ,
\end{equation}
where the coefficients
\begin{equation}\label{coef-hankel}
  \langle \nu , k\rangle=
  \frac{\Gamma\left(\frac 1 2 +\nu + k\right)}
  {k! \, \Gamma\left(\frac 1 2 +\nu - k\right)}
  =\langle -\nu , k\rangle
\end{equation}
are the Hankel symbols.

In particular, the quotient
\begin{equation}\label{JsobreJup}
      \frac{J_{\frac 1 2 -g}(\lambda)}{J_{g-\frac 1 2}(\lambda)}
      \sim e^{\pm i \pi \left(\frac 1 2 -g\right)}
      \frac{P(\frac 1 2 -g,\lambda)\mp i\,
      Q(\frac 1 2 -g,\lambda) }
      {P(g-\frac 1 2,\lambda) \mp i\,
      Q(g-\frac 1 2,\lambda) } = e^{\pm i \pi \left(\frac 1 2
      -g\right)}\, ,
\end{equation}
for $\Im(\lambda)>0$ and $\Im(\lambda)<0$ respectively, since
$P(\nu,z)$ and $Q(\nu,z)$ are even in $\nu$.

\bigskip

For the quotient of two Bessel functions we have
\begin{equation}\label{JsobreJup1}
      \frac{J_{\nu_1}(z)}{J_{\nu_2}(z)}
      \sim e^{\pm i \frac{\pi}{2} (\nu_1 - \nu_2)}\,
      \frac{P(\nu_1 ,z)\mp i\,
      Q(\nu_1 ,z) }
      {P(\nu_2,z) \mp i\,
      Q(\nu_2,z) }\, ,
\end{equation}
where the upper sign is valid for $\Im(\lambda)>0$, and the lower
one for $\Im(\lambda)<0$. The coefficients of these asymptotic
expansions can  be easily obtained, to any order, from Eq.\
(\ref{P+Q}),
\begin{equation}\label{asymp-cociente}\begin{array}{c}
    \displaystyle{\frac{P(\nu_1 ,z)\pm i\,
      Q(\nu_1 ,z) }
      {P(\nu_2,z) \pm i\,
      Q(\nu_2,z) }} \sim
      1 + \Big( \langle \nu_1 , 1\rangle -
     \langle \nu_2 , 1\rangle \Big) \,\left(\frac{\pm i}{2 z}\right)
      +  O\left(\frac 1 {z^2}\right)\, .
\end{array}
\end{equation}

\bigskip

Similarly, the derivative of the Bessel function has the following
asymptotic expansion \cite{A-S} for $|\arg z|<\pi$,
\begin{equation}\label{deriv-asymp}
  J'_\nu(z) \sim -\frac{2}{\sqrt{2 \pi  z}}
  \left\{
  R(\nu,z) \sin \chi(\nu,z) + S(\nu,z) \cos\chi(\nu,z)
  \right\}\, ,
\end{equation}
where
\begin{equation}\label{R}
  R(\nu,z) \sim \sum_{k=0}^\infty(-1)^k\,
  \frac{\nu^2 + (2k)^2-1/4}
  {\nu^2-(2k-1/2)^2}\,
  \frac{\langle\nu,2k\rangle
  }{\left(2 z\right)^{2 k}}\, ,
\end{equation}
and
\begin{equation}\label{S}
  S(\nu,z) \sim \sum_{k=0}^\infty(-1)^k \,
  \frac{\nu^2 + (2k+1)^2-1/4}
  {\nu^2-(2k+1-1/2)^2}\,
  \frac{\langle\nu,2k+1\rangle
  }{\left(2 z\right)^{2 k+1}}\, .
\end{equation}
Then,
\begin{equation}\label{Jprima}
  J'_\nu(z)\sim \mp i\,
  \frac{e^{\mp i z}\, e^{\pm i \pi \left(
  \frac \nu 2 + \frac 1 4 \right)}}{\sqrt{2\pi z}}
  \left\{ R(\nu,z)
  \mp i \,S(\nu,z) \right\}\, ,
\end{equation}
where the upper sign is valid for $\Im(\lambda)>0$, and the lower
one for $\Im(\lambda)<0$. We have also
\begin{equation}\label{RS}
  R(\nu,z) \pm i \,S(\nu,z)=
  P(\nu,z) \pm i \,Q(\nu,z)
  + T_\pm(\nu,z)\, ,
\end{equation}
with
\begin{equation}\label{T}
  T_\pm(\nu,z)\sim
  \sum_{k=1}^\infty
  (2k-1)\langle\nu,k-1\rangle
  \left(\frac{\pm i}{2z}\right)^{k}\, .
\end{equation}

Therefore, we get
\begin{equation}\label{JprimasobreJ}
    \frac{J'_\nu(z)}{J_\nu(z)}
  \sim \mp  i \left\{
  1+ \frac{T_\mp(\nu,z)}{P(\nu,z)\mp i Q(\nu,z)}\right\}\, ,
\end{equation}
where the upper sign is valid for $\Im(\lambda)>0$, and the lower
one for $\Im(\lambda)<0$. The coefficients of the asymptotic
expansion in the right hand side of Eq.\ (\ref{JprimasobreJ}) can
be easily obtained from Eq.\ (\ref{P+Q}) and (\ref{T}),
\begin{equation}\label{TsobrePQ}\begin{array}{c}
    \displaystyle{
   \frac{T_\pm(\nu,z)}{P(\nu,z)\pm i Q(\nu,z)}
  = \left(\frac{\pm i}{2 z}\right)
  + O\left(\frac 1 {z^2}\right)
  }
\end{array}
\end{equation}

Finally, since the Hankel symbols are even in $\nu$ (see Eq.\
(\ref{coef-hankel})), from Eq.\ (\ref{P+Q}), (\ref{T}) and
(\ref{JprimasobreJ}) we have
\begin{equation}\label{JprimasobreJasymp}
  \frac{J'_\nu(z)}{J_\nu(z)}
  \sim \frac{J'_{-\nu}(z)}{J_{-\nu}(z)}\, .
\end{equation}

\bigskip

\end{document}